\documentstyle[prb,aps,twocolumn]{revtex}

\begin{document}

\twocolumn[
\hsize\textwidth\columnwidth\hsize\csname@twocolumnfalse\endcsname

\draft
\title{Electrical transport studies of quench condensed Bi films\\ 
at the initial stage of film growth:
Structural transition and the \\possible formation of electron droplets}
\author{M. M. Rosario and Y. Liu}
\address{Department of Physics, The Pennsylvania State University,
  University Park, PA 16802}
\date{\today}
\maketitle

\begin{abstract}

The electrical transport properties of amorphous Bi films prepared by
sequential quench deposition have been studied {\it in situ}. A
superconductor-insulator (S-I) transition was observed as the film was
made increasingly thicker, consistent with previous studies. Unexpected
behavior was found at the initial stage of film growth, a regime not
explored in detail prior to the present work. As the temperature was
lowered, a positive temperature coefficient of resistance (d$R$/d$T > 0$)
emerged, with the resistance reaching a minimum before the d$R$/d$T$
became negative again.  This behavior was accompanied by a non-linear and
asymmetric $I-V$ characteristic.  As the film became thicker, conventional
variable-range hopping (VRH) was recovered.  We attribute the observed
crossover in the electrical transport properties to an amorphous to
granular structural transition. The positive d$R$/d$T$ found in the
amorphous phase of Bi formed at the initial stage of film growth was
qualitatively explained by the formation of metallic droplets within the
electron glass.

\end{abstract}


]

\narrowtext

\section{Introduction}
\label{sec:intro}

Quench condensed films, prepared by depositing metal vapor directly onto
substrates held at liquid helium temperatures,\cite{shalnikov38} have
played important role in elucidating the effects of interactions,
disorder, and geometrical constraints on the properties of an electronic
system. Films made in this manner are conducting at thicknesses as low as
$d\approx 10$\AA,\cite{komnik82} making it a two-dimensional (2D)
disordered system.  Quench condensed films have been used in numerous
experimental studies of superconductivity,\cite{sc}
localization,\cite{loc} and their interplay. More recently, it has played
an important role in the study of quantum phase transitions, such as the
2D superconductor-insulator (S-I) and metal-insulator (M-I)
transitions.\cite{qpt}

Physics in a strongly disordered system is dominated by electron-electron
interactions. These interactions result in the formation of a Coulomb gap,
namely a reduction in the single-particle density of states (DOS) near the
Fermi energy,\cite{pollak70} leading to a modification of
variable-range-hopping (VRH) conduction.\cite{efros75and76} The
temperature dependence of electrical conductivity in the VRH regime is
given by $\sigma (T) = \sigma _{\text o}{\text {exp}}[-(T_{\text
o}/T)^a]$, where $\sigma $ is the electrical conductivity, $T$ is
temperature, $\sigma _{\rm o}$ and $T_{\rm o}$ are constants, and $a=1/3,
1/4$ in non-interacting 2D and 3D systems, respectively. The introduction
of Coulomb interactions results in $a=1/2$, regardless of the
dimensionality, with the constant $T_{\rm o}$ given by $T_{\rm o} = {\rm
e}^2/({\rm k}_{\rm b} \epsilon \xi _{\rm L})$, where $\epsilon $ is the
dielectric constant and $\xi _{\rm L}$ is the localization length.

The Coulomb gap problem can be mapped onto an Ising spin glass model
with $1/r$ antiferromagnetic interactions in a random field, suggesting
that a finite-temperature glass transition may be
present.\cite{davies82,davies84,xue88,ruiz93,diaz-sanchez00}  An order
parameter, similar to the Edwards-Anderson order parameter used for spin
glass systems, was proposed for the electron glass in an attempt to find
a glass transition. Though a non-zero value of the order parameter was
found, critical behavior was absent in the numerical studies. 
Subsequent work\cite{xue88} revealed that a glass transition is not
present at non-zero temperatures, similar to 2D Ising spin
glasses.\cite{binder86}

Experimental studies of the electron glass were carried out originally on
lightly doped semiconductors.\cite{semiconductors} Recently,
field-effect experiments on strongly disordered
systems,\cite{InOx,adkins84} including quench condensed
films,\cite{m-arizala97and98} have revealed glassy behavior (but not
critical behavior associated with a glass transition). In particular,
effects such as slow relaxation, hysteresis, memory and aging are
documented in quench condensed Bi and Pb.\cite{m-arizala97and98} In
addition, an observed minimum in the conductance as a function of gate
voltage was proposed to be related to the opening of the Coulomb gap in
the DOS.\cite{yu99} Interestingly, though the conductance for the films
was found to fit the general hopping form, the value for $a$ obtained from
the fitting was close to 0.8,\cite{m-arizala97and98,markovic00} a
dependence not given by any VRH theory.

Recently, {\it in situ} scanning tunneling microscopy (STM) studies on
granular quench condensed films of Au and Pb revealed an unexpected
scenario for film growth.\cite{ekinci98and99} It appears that a uniform,
homogeneously disordered phase formed during the initial depositions. This
precursor phase was found to avalanche into an islanded structure as the
thickness reached a critical value. Results obtained on both Au and Pb
films indicate that the observed structural evolution may be general,
independent of substrate material and film preparation procedure.  A
similar amorphous precursor phase was found in {\it in situ}
Raman-scattering measurements on Bi films deposited onto carbon substrates
at 110K.\cite{lannin91} The Raman spectra indicated a distinct local
structural transition from amorphous-like to nanocrystalline clusters as
the film thickness increased across a critical thickness of $\sim 8$\AA.

These uniform, amorphous films should also form at the initial stage of
film growth in Bi films quench condensed at liquid helium temperatures,
even though the local amorphous to crystalline structural transformation
may be incomplete because of the reduced substrate temperature. The
electrical transport properties of these films at the initial stage of
film growth, which have never been studied as the regular sized films are
too resistive to measure, are of great interest. Transport measurements on
films of macroscopic size, taken simultaneously with the {\it in situ} STM
studies, showed that the onset of conductivity occurred when the films
were seen to have two nearly complete layers of grains, having formed
electrically connected multigrain clusters.  The undetectable electrical
conductivity of the initial amorphous precursor phase on a macroscopic
scale suggests either electron localization at an atomic scale because of
the strong disorder,\cite{danilov} or structural discontinuity at a large
length scale even though electrons within each connected amorphous region
are not necessarily strongly localized.  The prevailing view appears to
favor the former scenario.

We have studied the electrical transport properties of quench condensed
Bi films in this precursor regime.  While conventional glassy behavior
was found in these thinnest films, an unexpected positive temperature
coefficient of resistance (d$R$/d$T >0$), accompanied by a non-linear
and asymmetric current-voltage ($I-V$) characteristic, was observed in
the initial stage of film growth.  As more Bi was deposited, this state
was found to enter a more familiar state characterized by VRH with a
Coulomb gap.  This crossover in electrical transport properties appeared
to coincide with an amorphous to granular structural transition in
ultrathin Bi reported previously.\cite{ekinci98and99} The positive
d$R$/d$T$ found in the initial films is proposed to be a result of the
formation of electron droplets within an electron glass state.

\section{Sample Preparation and Measurements}
\label{sec:sampleprep}

Films were prepared by quench deposition in a ${}^3{\text {He}}$
cryostat.\cite{rosario00} A glazed alumina (amorphous Al$_2$O$_3$)
substrate, held at liquid helium temperatures during the evaporation of
Bi, was used. An amorphous Ge underlayer, previously used to help grow an
electrically homogeneous film, was {\it not} evaporated onto the substrate
prior to the Bi evaporation. Current biased d.c. electrical transport
measurements were taken {\it in situ} after each evaporation. A set of
films was created by sequential deposition. To ensure that any possible
annealing was avoided, all films were kept at temperatures lower than 10
K, and bias currents of less than 1 nA were used.

A major difficulty in studying the electrical transport properties of the
films at the initial stage of growth is the extremely high resistivity,
resulting in an unmeasurable resistance for samples of conventional size.  
This difficulty can be circumvented by preparing films of extremely short
length but relatively large width, which can be conveniently done using a
2-point probe configuration.  An undesired consequence of this approach is
the somewhat uncontrolled contact resistance associated with 2-point
measurements.  We have addressed this experimental issue in two ways.  
First, we made an effort to design the sample configuration that would
minimize the contact resistance.  Second, we studied samples of different
lengths, so that the contact resistance can be inferred from a scaling
analysis.

In the present work, 100 \AA \enskip thick Au measurement leads were
pre-patterned onto the substrate. Films were deposited across a narrow gap
between two Au contacts and atop the entire Au electrode.  Since the Au
film is significantly more conductive than the initial Bi films, the
measured resistance is dominated by the resistance of the Bi film in the
gap and the contact resistance.  For the latter, the following steps were
taken to minimize its magnitude.  The gap was prepared using a shadow
mask, resulting in a smooth and gradual edges, as revealed by atomic force
microscope (AFM) study.  The combination of the gradual slope and the good
conductivity of the Au thin film is expected to minimize the contact
resistance, as appeared to be the case experimentally (see below). We
studied two film sequences of lengths 2.5 $\mu $m and 35 $\mu $m,
respectively, both 0.5 mm in width.  A schematic of the sample
configuration is shown in the inset of Fig. \ref{Vvst}.

Another experimental difficulty in measuring films at the initial stage of
film growth is the exceedingly long equilibrium times, due to both the
large RC constant and intrinsic glassy behavior.  We monitored the time
response of the voltage for all films which exhibited long settling times.
All values shown are the limiting values of the voltage rise after
adequate time was allowed for the measurement to stabilize.  The glassy
state previously observed in strongly disordered systems are often
characterized by extremely long, non-exponential relaxation
times.\cite{adkins84,m-arizala97and98} Similar behavior was observed in
the present set of films in the region where R $> 10^7 \Omega $
(corresponding to sheet resistances of R$_{\Box } > 10^9$), as
shown in Fig.\ref{Vvst}.  The voltage response can be fit to a
stretched exponential of the form $V(t) = V_{\rm o}(1 - \exp [-(t/ \tau
)^\alpha ])$, where $V_{\rm o}$ is the limiting value of the voltage,
$\tau $ the time constant, and $\alpha $ the exponential constant. The
time constants obtained were twice as large as expected from the RC
constant of the measurement system.

\section{Experimental Results}
\label{sec:expresults}

In addition to long relaxation times, films at the initial stage were
found to have a non-linear, asymmetric $I$-$V$ characteristic, as seen in
Fig. \ref{VvsI}. The asymmetry and non-linearity were more pronounced at
lower temperatures and smoothly evolved to simple ohmic behavior as
temperature was increased.  Non-linearity, in itself, is not unexpected in
systems dominated by VRH conduction. However, an analysis of the data
using theoretically predicted dependences of conductance on electric field
in the VRH regime\cite{VRHiv} was inconclusive, as a satisfactory fit was
not found. The asymmetry about zero current bias in both magnitude and
curvature seen in the $I$-$V$ curves has not been theoretically predicted,
but may be understood qualitatively in a model proposed here (see below).

Given the nature of the $I$-$V$ characteristic, d.c. resistances were
calculated by measuring the voltage across the sample at a fixed current
bias of +20 pA.  The voltage at zero bias current is subtracted from this
value to account for offset voltages, a common practice in electrical
transport measurements. The resistance characteristic at negative current
bias is qualitatively similar to that measured with a positive bias.

The resistance as a function of temperature R($T$) of the 2.5$\mu $m long
film, at a thickness $d$ of 12.5\AA, is given in Fig. \ref{RvsT}.  Inset
a) shows a Zabrodskii plot\cite{zabrodskii84} of the data, where the
function $\ln[-d(\ln R)/d(\ln T)]$ is plotted against $\ln T$.  If the
resistance is described by the hopping conduction formula, $R= R _{\text
o}{\text {exp}}[(T_{\text o}/T)^a]$, we expect a straight line in such a
plot. The hopping exponent $a$ would correspond to the slope of the line.
A least squares line fit yielded a value $a=0.49\pm 0.07$, consistent with
VRH with Coulomb interactions.  A fit of the data specifically to this
form is given in inset b). The localization length is estimated to be $\xi
_{\rm L} \approx 80$\AA, assuming that the dielectric constant of the film
is that of the substrate. Similar fits to simple activated behavior
($a=1$) and non-interacting VRH conduction ($a=1/3, 1/4$) were
unsatisfactory.  The temperature dependence of the conductivity and the
slow voltage response indicate that these films can be characterized as
being in an electron glass state.

As indicated in Fig. \ref{RvsT} inset b), the resistance deviated from VRH
(with interactions) at low temperature.  The temperature at which the
deviation began was close to the temperature at which the $I$-$V$
characteristic exhibited asymmetric and non-linear behavior.  In thinner
films, the temperature above which VRH was observed increased, with the
resistance deviation persisting to increasingly higher temperatures. In
the thinnest films, standard hopping conduction was no longer observed
within the temperature range studied ($< 10$ K).  In addition, a
non-linear and asymmetric $I$-$V$ characteristic was present over the
whole temperature range.

Remarkably, in these thinnest films, the low temperature deviation
developed into a positive temperature coefficient of resistance, i.e.  
d$R$/d$T > 0$. Fig. \ref{RvsT2}a-d shows the R($T$) of films in this
regime in detail. The positive d$R$/d$T$ did not extend down to $T=0$, but
instead reached a resistance minimum and recovered insulating behavior in
the limit of zero temperature.  The resistance at intermediate
temperatures showed an almost linear temperature dependence.  In contrast,
the low temperature insulating behavior was found to rise dramatically
with decreasing temperature.  The precise temperature dependences in these
two regimes were difficult to determine due to the small temperature range
available for data fitting.

An examination of behavior in films of different thicknesses revealed
that the positive d$R$/d$T$ persisted to higher temperatures with
increasing disorder, as seen in Figs. \ref{RvsT2}a-d. In contrast, the
resistance minimum occurred at $T\approx 2.5-3$K, for all films which
exhibited a positive d$R$/d$T$, regardless of film thickness. This
feature disappeared at $d =  12.0$\AA. 

The evolution of the film resistance with respect to film thickness is
shown in Fig. \ref{bigRvsT}.  For $d \geq 12.0$\AA, a striking departure
from the behavior of the thinnest films was seen in the resistance
characteristic.  Although a non-linear and asymmetric $I-V$ characteristic
was still observed at the lowest temperatures, a resistance minimum was no
longer found. The films exhibited an insulating character over the whole
temperature range ($T < 8$K).  For films with $d > 12.5$\AA, the $I$-$V$
curves displayed ohmic behavior over the whole temperature ($T < 8$K) and
current range ($I < 100$ pA).  As more material was added the resistance
was observed to enter a region where the resistance could not be fitted to
any theoretically predicted dependence.

As the film was made thicker, familiar behavior for granular ultrathin
films of superconducting metals, including a superconductor-insulator
(S-I) transition was found, similar to what has been previously
observed.\cite{jaeger89} On the insulating side of the transition, the
resistance displayed hopping conduction characteristics at higher
temperatures, but leveled off at low temperatures.  The level-off
phenomenon has been widely observed in ultrathin insulating
films.\cite{jaeger89,wu94} Further increases in thickness leads to a
weakly localized regime where the resistance has a $\ln T$ dependence.

The kink in the R($T$), below which a more rapid rise in resistance was
seen, has been attributed to the opening of the superconducting energy gap
in individual grains at a transition temperature $T_{\rm
{co}}$.\cite{haviland89} As a result, the activation energy for single
electron hops between grains is the sum of the charging and the
superconducting pairing energy.\cite{haviland89,dynes78,adkins80} As more
material was deposited, the kink developed into a resistance minimum, a
well known phenomenon of quasi-reentrant
superconductivity.\cite{haviland89,orr85} In this case, although
individual grains become superconducting with zero resistance which causes
a resistance drop at T$=T_{\rm {co}}$, the phases of these grains do not
establish a long-range coherence.\cite{simanek79and82} Quantum
fluctuations drive the film insulating at lower
temperatures.\cite{fisher86} The observed re-entrant behavior, which has
become the hallmark for granular superconducting films, indicates that
the Bi films prepared in the present study were granular as well.

As seen in Fig. \ref{bigRvsT}, the evolution of the resistance as a
function of film thickness is remarkably systematic: in the thinnest
films, the positive temperature coefficient of resistance becomes less
distinct in relatively thicker films; at intermediate thicknesses the
films develop standard VRH behavior at higher temperatures, as well as
with increasing thickness; finally, further increases in film thickness
result in an S-I transition, consistent with previous studies.  
Qualitatively similar behavior was also observed in studies of the 35 $\mu
m$-long film.

It is important to examine the issue concerning the contact resistance
between the Bi and the Au contacts. A comparison of resistances of the
co-deposited 2.5 $\mu$m- and 35 $\mu$m-long films, as well as values of
contact resistance, are shown in Table 1.  Films thinner than 11.0\AA
\enskip have contact resistances that are neglible compared to the film
resistance.  For films thicker than 15.5\AA, the contact resistance is
found to be a significant fraction of the film resistance.  However, as
noted above, the behavior is strikingly similar to what was previously
observed in this system using a 4-point configuration.  This indicates
that, despite the 2-point configuration, the intrinsic nature of the film
is reflected in the measurement.  This is possible if the contact
resistance, though large, is temperature independent.  Finally, films of
intermediate thicknesses were found to not scale, and a value for the
contact resistance could not determined; this issue will be discussed in
the following section.

\section{Amorphous to granular structural transition}
\label{sec:structuraltransition}

The intriguing crossover of the resistance characteristic with respect to
film thickness observed at around $12.0-14.0$\AA \enskip in quench
condensed Bi films has raised the question concerning its physical origin.
Below we argue that this crossover may be related to film growth models
advanced by {\it in situ} Raman studies on quench condensed
Bi,\cite{lannin91} and STM results obtained on quench condensed Au and Pb
films.\cite{ekinci98and99}

The abrupt disappearance of the resistance minimum in the thinnest films,
which marked a qualitative change in the resistance characteristic,
appears to point to a scenario similar to the avalanche model proposed to
explain the {\it in situ} STM observations.  Fig. \ref{rvsd} shows the
sheet resistance, $R_{\Box }$, at T $=8$K for the entire set of films.
Within the range 11.5\AA$ < d < 14.0$\AA, $R_{\Box }$ decreased
particularly rapidly with increasing thickness, indicating that the films
underwent a qualitative change in this thickness range.

The behavior seen in Fig. \ref{rvsd} indicates the existence of two
distinctly different regimes, with a rapid crossover region observed over
a change in film thickness of only around 2\AA. In these two regions, the
thickness dependences of $R_{\Box }$ were similar, exhibiting a relatively
slow change with increasing thickness in comparison with the transition
region, though the values of $R_{\Box }$ in the two limits ($d < 11.5$\AA
\enskip and $d > 14.0$\AA) differ by more than 5 orders of magnitude.
Interestingly, at $t = 13.0\AA$, a thickness corresponding to the middle
of the transition region, the film resistance does not follow any
theoretically predicted dependence and, instead, appears as a boundary
between the high resistance and low resistance regions.  The significant
difference in magnitude of the resistivity, the sharp transition, as well
as the distinct difference in character between the two regimes may be
explained by a change in the film's structure as the thickness is
increased.

Such a change is further supported by the lack of scaling between the 2.5
$\mu $m- and 35 $\mu $m-long film in and near the transition region (see
Table 1).  As the films undergo the transition, a characteristic length
determines the films behavior.  Films whose size is larger than this
length should scale, whereas those smaller should not. The rapidly
changing resistance found in this thickness range suggests that this
length is strongly thickness dependent. As the films moved farther into
the low resistance region, they were again found to scale, indicating the
end of the transition, and the emergence of a stable film structure.

Within this structural transition scenario, the film develops from a
uniform to granular film at a critical thickness $d \approx 11.5$\AA.  
Assuming that the resistance in the granular film after the transition was
dominated by intergrain electron hopping, further material depositions
serves to decrease the tunnel barrier strength between grains, increasing
the conductivity.  Therefore, the resistance of the films would depend on
thickness more sensitively in just-formed granular films, and is less
sensitive in the relatively thicker films after a path of grains across
the film is already established, consistent with experimental observation.

In addition, within each individual grain, a local amorphous to
crystalline structural transformation occurs, similar to that observed in
{\it in situ} Raman-scattering measurements in Bi films deposited at 110K.  
In that study, the transition from an amorphous, disordered structure to a
rhombohedral phase was found between 7.0\AA \enskip and 8.5\AA.  The
presence of such crystalline-like grains is supported by an observed
$T_{\rm {co}}$ of 2.5 -- 3.0K near the S-I transition.  Similar values of
$T_{\rm {co}}$ were found in nanocrystalline Bi
clusters.\cite{nanocrystallineBi} However, given the lower deposition
temperatures used in the present study and, therefore, the lower thermal
energy available for Bi atom diffusion, a larger critical thickness for
the local amorphous to crystalline transition ($d \approx 11.5\AA$) would
be natural. Furthermore, the local transition in the present films could
be incomplete for the same reason, resulting in grains that are between
amorphous and crystalline in nature.  We refer to this phase as
quasi-crytalline. The thickness dependence of $T_{\rm {co}}$, {\it i.e.}
that $T_{\rm {co}}$ increases with increasing thickness to a maximum value
of 6.5 K, provides further support that the grains after the structural
transition retain a substantially amorphous character.

The proposed phase diagram for this transition is shown in the inset of
Fig. \ref{rvsd}.  As temperature increases, an additional, previously
studied transition to a nanocrystalline phase\cite{komnik82} is expected
(not indicated in the schematic).

\section{Possible droplet formation\\in the electron glass}
\label{sec:droplet}

Another question raised by the present work is the nature of the
electrical transport properties of the amorphous films found at the
initial stage of film growth.  It is remarkable that the temperature at
which the resistance starts to rise in these films coincides with the
$T_{\rm {co}}$ of the locally superconducting, granular films near the S-I
transition, as seen in Fig. \ref{bigRvsT}.  This observation can be
understood if we assume that in the thinnest films, which should be
homogeneously disordered over a large area, there may be locally
superconducting regions (superconducting droplets).  The resistance
increase at low temperature would then be related to the opening of the
energy gap in the superconducting droplets, similar to the granular
superconducting film.  The question is whether these droplets could indeed
form in these strongly insulating films.

We note that a droplet state was proposed in a theoretical study on the 2D
interacting, but highly mobile, electron system.\cite{shi99he98} Motivated
by the recent observations of the 2D MIT,\cite {kravchenko00} it was
suggested that such a system is unstable against phase separation into a
high-density Fermi gas (``liquid'' phase) and a low density insulating
Wigner crystal (``gas'' phase).  The formation of a two-phase coexistence
region, a droplet state, was shown to be favored by increasing potential
disorder.  Similarly, droplets might have begun to form in these amorphous
films at a characteristic temperature, growing in size as temperature
decreases.  The resistance is determined by single electron hops between
droplets. As the distance between droplets decrease with decreasing
temperature, as does the resistance.  This results in the observed
deviation from VRH as well as the positive d$R$/d$T$.

The formation of these droplets minimizes the Coulomb repulsion energy of
the system in order to be stable.  However, the entropy of the system
decreases as a result.  At lower temperatures, the energy change is likely
sufficient to compensate for the loss of entropy. At higher temperatures,
the entropy term becomes more important and droplet formation may be less
favored, resulting in a "gas" (uniform electron glass) state with
insulating behavior, consistent with the thickness dependence of the onset
of the positive d$R$/d$T$.

This picture can also qualitatively account for the asymmetry in the
$I$-$V$ characteristic.  Nucleation sites for the droplets are highly
dependent on the potential disorder landscape present in the system, which
is, in turn, dependent on the applied electric field.  Taking into account
that the films' left and right interfaces are not identical, the effective
applied field need not be symmetric with respect to field direction.  It
follows that the spatial distribution of nucleation sites are unidentical,
and the lowest resistance path is not expected to be the same for
different field polarities.

The formation of such droplets may be natural if we consider the
one-to-one correspondence of the Coulomb glass to the Ising spin glass.  
A phenomenological scaling theory of the ordered phase of short-range
Ising spin glasses was developed in terms of droplet
excitations.\cite{fisher88} In this model, low lying droplet excitations,
consisting of connected clusters of spins reversed from the ground state,
dominate the physics of the spin glass. A similar phenomena would not be
unexpected in the electron glass.

\section{Summary}
\label{sec:summary}

We have measured the electrical transport properties of quench condensed
Bi films deep in the insulating regime which has never been explored prior
to the present work.  Unexpected behavior was seen in the thinnest films.  
These anomalies, specifically, a state characterized by a positive
d$R$/d$T$, accompanied by a non-linear and asymmetric $I$-$V$
characteristic, were present only in the initial stage of film growth. As
more material is deposited, this unconventional state abruptly disappears.
After a transition regime, the films were found to undergo an S-I
transition observed previously.

The evolution of the transport properties with respect to film thickness
at the initial stage of film growth is attributed to a structural
transition from an amorphous to granular structure in these thinnest
quench condensed Bi films. The positive d$R$/d$T$ found in the amorphous,
precursor layers were proposed to result from the formation of high
electron density droplets within the insulating electron glass background.  
Though this model can qualitatively account for the features seen in the
conduction, the nature of the amorphous precursor phase needs to be
further studied, both experimentally and theoretically.

\section{Acknowledgements}
\label{sec:acknowledgements}

We acknowledge useful discussions with X. Xie and W. Wu, and technical
assistance from Yu. Zadorozhny.  This work is supported by the National
Science Foundation through Grant DMR-9702661.



\begin{figure}

  \caption{The time response of the voltage to an applied current bias
of +20pA at $T =0.60$K.  The solid line represents a fit to $V(t) =
V_{\rm o}(1 - \exp [-(t/\tau )^\alpha ])$ with $V_{\rm o}=0.22$V, $\tau
=80$s, and $\alpha =0.8$. A schematic of the sample configuration is
shown in the inset.}
  \label{Vvst}

\end{figure}

\begin{figure}
  \caption{$I$-$V$ characteristic of quench condensed Bi at a thickness of
12.5\AA.  The voltages given are the limiting values,
with the offset voltage for $I=0$ subtracted for each current. The
temperatures are 0.60K, 0.80K, 0.90K, 1.20K, 1.40K, 1.80K,
2.00K, 2.20K, 2.40K, 2.80K and 4.00K.  The first fully linear curve was
found at $T\approx 2.2$K.  All resistance measurements taken deep in the
insulating regime were made using a +20pA current bias.}
  \label{VvsI}
\end{figure}

\begin{figure}
  \caption{The resistance as a function of temperature R($T$) for a
12.5\AA \enskip thick film, using a +20pA current bias. A Zabrodskii
plot of the data, shown in inset a), indicates that a hopping conduction
exponent of $a=1/2$ best describes the resistance at higher
temperatures.  This dependence is best seen on a semilog plot of the R
versus $\sqrt T$ as given in inset b). The solid line is a fit to VRH
with Coulomb interactions. The resistance deviated from this behavior at
$T \approx 2.2$K.}
  \label{RvsT}
\end{figure}

\begin{figure}
  \caption {The resistance characteristic at film thicknesses of
a) 10.0\AA, b) 10.5\AA, c) 11.0\AA, and d) 11.5\AA.  The lines are a
guide to the eye.}
  \label{RvsT2}
\end{figure}

\begin{figure}
  \caption{Thickness dependence of the R($T$) for the 2.5$\mu $m long
film.  The dashed line denotes the temperature at which the
resistance minima were found in the high-disorder limit.  This
temperature is seen to be close to the local $T_{\rm c}$ near
the S-I transition.}
  \label{bigRvsT}
\end{figure}

\begin{figure}
  \caption{A semi-log plot of the sheet resistance ${\text R}_{\Box }$ at
$T = 8$K as a function of film thickness.  The inset shows a proposed
phase diagram for the amorphous to granular structural transition in
quench condensed Bi films at the initial stage of film growth.}
  \label{rvsd}
\end{figure}

\begin{table}
  \caption{Contact resistance, $R_{\text c}$, at various film thicknesses.
Resistance values of the co-evaporated $35 \mu $m ($R_{\text {long}}$) and
$2.5 \mu$m ($R_{\text {short}}$) films were compared at $T = 8$K. For
films of thickness $11.0\AA \enskip < d < 15.5\AA$, the film resistances
were found to not scale and a value of the contact resistance could not be
determined.}
  \label{table1}
  \begin{tabular}{dlll}
  thickness (\AA)               & $R_{\text  {long}} (\Omega )$ &
  $R_{\text {short}} (\Omega )$ & $R_{\text c}(\Omega )$\\
  \tableline
  9.5  & 2.50$\pm $0.05$\cdot 10^{11}$ & 2.00$\pm $0.05$\cdot 10^{10}$ 
	&2.3$\pm $0.9$\cdot 10^9$\\
  11.0 & 3.70$\pm $0.05$\cdot 10^{10}$ & 2.7$\pm $0.1$\cdot 10^9$   
	&9$\pm $6$\cdot 10^7$\\
  12.0 & 1.36$\pm $0.05$\cdot 10^{10}$ & 8.4$\pm $0.1$\cdot 10^7$    &\\
  12.5 & 5.8$\pm $0.1$\cdot 10^9$      & 7.0$\pm $0.5$\cdot 10^6$    &\\
  13.0 & 9.8$\pm $0.2$\cdot 10^8$      & 5.2$\pm $0.1$\cdot 10^5$    &\\
  14.0 & 5.60$\pm $0.05$\cdot 10^6$    & 2.35$\pm $0.05$\cdot 10^4$    &\\
  14.5 & 5.8$\pm $0.5$\cdot 10^5$      & 1.4$\pm $0.1$\cdot 10^4$    &\\
  15.5 & 2.3$\pm $0.2$\cdot 10^5$      & 6.6$\pm $0.1$\cdot 10^4$    &\\
  17.5 & 2.91$\pm $0.05$\cdot 10^4$    & 2.59$\pm $0.05$\cdot 10^3$    
	& 5.4$\pm $0.9$\cdot 10^2$\\
  20.5 & 1.80$\pm $0.03$\cdot 10^3$    & 7.00$\pm $0.02$\cdot 10^2$   
	&6.1$\pm $0.2$\cdot 10^2$\\
  22.5 & 9.73$\pm $0.05$\cdot 10^2$    & 4.17$\pm $0.05$\cdot 10^2$    
	&3.74$\pm $0.06$\cdot 10^2$\\
  \end{tabular}
  \end{table}


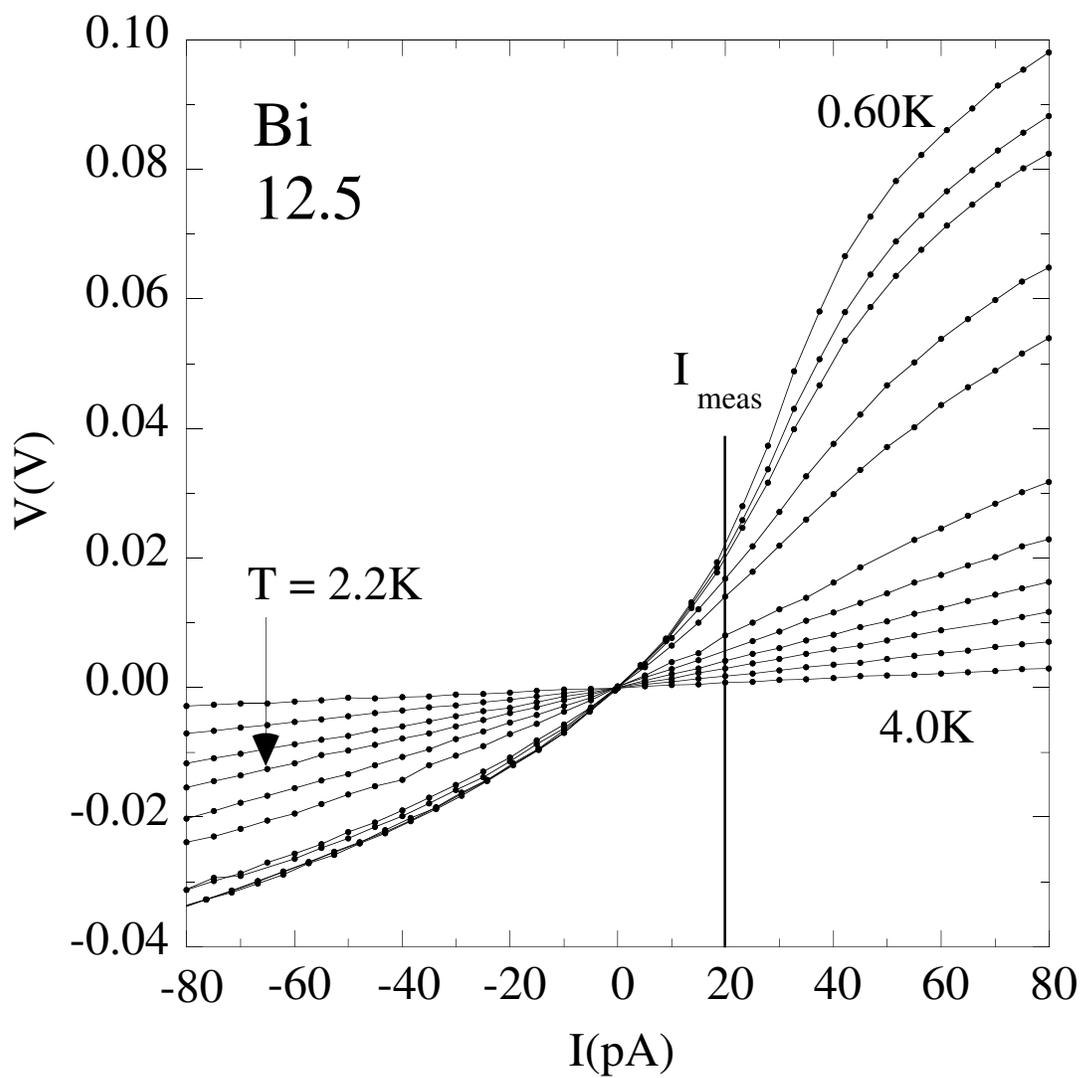
\end{document}